\documentclass[fleqn,usenatbib]{mnras}

\usepackage{newtxtext,newtxmath}

\usepackage[T1]{fontenc}

\DeclareRobustCommand{\VAN}[3]{#2}
\let\VANthebibliography\thebibliography
\def\thebibliography{\DeclareRobustCommand{\VAN}[3]{##3}\VANthebibliography}

\usepackage{graphicx}	
\usepackage{amsmath}	

\newcommand{\ms}{\,ms$^{-1}$}	
\newcommand{\mjup}{\,$M_{\rm J}$}	
\newcommand{\rjup}{\,$R_{\rm J}$}

\newcommand{\tess}{{\it TESS}}

\newcommand{\gaia}{{\it Gaia}}
\newcommand{\jwst}{{\it JWST}}
\newcommand{\ngts}{{NGTS}}
\newcommand{\wasp}{{WASP}}
\newcommand{\coralie}{{CORALIE}}
\newcommand{\harps}{{HARPS}}
\newcommand{\feros}{{FEROS}}
\newcommand{\chiron}{{CHIRON}}
\newcommand{\pfs}{{PFS}}

\newcommand{\teff}{{T$_{\rm eff}$}}
\newcommand{\logg}{{$\log$ g}}
\newcommand{\feh}{[Fe/H]}
\newcommand{\vsini}{$V\sin i$}
\newcommand{\systemt}{{\rm TOI-2447}}
\newcommand{\systemtic}{{\rm TIC-1167538}}
\newcommand{\systemtb}{{\rm TOI-2447\,b}}

\newcommand{\systemn}{{\rm NGTS-29}}
\newcommand{\systemnb}{{\rm NGTS-29\,b}}
\newcommand{\system}{{\rm TOI-2447}}
\newcommand{\systemb}{{\rm TOI-2447\,b}}

\title[Discovery of TOI-2447\,b $\slash$ NGTS-29\,b: a 69-day Saturn]{TOI-2447\,b $\slash$ NGTS-29\,b: a 69-day Saturn around a Solar analogue}

\author[S. Gill et al.]{\parbox{\textwidth}{\Large
Samuel Gill$^{1,2}$,
Daniel Bayliss$^{1,2}$,
Sol\`ene Ulmer-Moll$^{3,39}$,
Peter J. Wheatley$^{1,2}$,
Rafael Brahm$^{4,5,6}$,
David R. Anderson$^{1,2}$,
David Armstrong$^{1,2}$,
Ioannis Apergis$^{1,2}$,
Douglas R. Alves$^{7,8}$,
Matthew R. Burleigh$^{9}$,
R.P. Butler$^{10}$,
Fran\c{c}ois Bouchy$^{3}$,
Matthew P. Battley$^{3}$,
Edward M. Bryant$^{11}$,
Allyson Bieryla$^{12}$,
Jeffrey D. Crane$^{13}$,
Karen A.\ Collins$^{12}$,
Sarah L. Casewell$^{9}$,
Ilaria Carleo$^{14}$,
Alastair B. Claringbold$^{1,2}$,
Paul A. Dalba$^{15}$,
Diana Dragomir$^{16}$,
Philipp Eigm\"uller$^{17}$,
Jan Eberhardt$^{18}$,
Michael Fausnaugh$^{19}$,
Maximilian N. G{\"u}nther$^{20}$,
Nolan Grieves$^{3}$,
Michael R. Goad$^{9}$,
Edward Gillen$^{21,22}$,
Janis Hagelberg$^{3}$,
Melissa Hobson$^{18,5}$,
Christina Hedges$^{23}$,
Beth A. Henderson$^{9}$,
Faith Hawthorn$^{1,2}$,
Thomas Henning$^{18}$,
Mat\'ias I. Jones$^{24}$,
Andrés Jordán$^{4,5,6}$,
James S. Jenkins$^{25,8}$,
Michelle Kunimoto$^{19}$,
Andreas F. Krenn$^{3,26}$,
Alicia Kendall$^{9}$,
Monika Lendl$^{3}$,
James McCormac$^{1,2}$,
Maximiliano Moyano$^{24}$,
Pascal Torres-Miranda$^{5,27}$,
Louise D. Nielsen$^{28}$,
Ares Osborn$^{1,2}$,
Jon Otegi$^{3}$,
Hugh Osborn$^{19}$,
Samuel N. Quinn$^{12}$,
Joseph E. Rodriguez$^{29}$,
Gavin Ramsay$^{30}$,
Martin Schlecker$^{31}$,
Stephen A. Shectman$^{13}$,
Sara Seager$^{19,32,33}$,
Rosanna H. Tilbrook$^{9}$,
Trifon Trifonov$^{18,38}$,
Johanna K. Teske$^{10}$,
Stephane Udry$^{3}$,
Jose I. Vines$^{24}$,
Richard R. West$^{1,2}$,
Bill Wohler$^{34}$,
Joshua N. Winn$^{35}$,
Sharon X. Wang$^{36}$,
George Zhou$^{37}$,
Tafadzwa Zivave$^{1,2}$
}
\vspace{0.2cm}
\\
\parbox{\textwidth}{
The authors' affiliations are shown in the Appendix.\\
*E-mail: samuel.gill@warwick.ac.uk}\vspace{-0.3cm}}

\date{Accepted XXX. Received YYY; in original form ZZZ}

\pubyear{2022}

\begin{document}
\label{firstpage}
\pagerange{\pageref{firstpage}--\pageref{lastpage}}
\maketitle

\begin{abstract}
Discovering transiting exoplanets with relatively long orbital periods ($>$10\,days) is crucial 
to 
facilitate the study of cool exoplanet atmospheres ($T_{\rm eq} < 700 K$) and to understand exoplanet formation and inward migration further out than typical transiting exoplanets.  In order to discover these longer period transiting exoplanets, long-term photometric and radial velocity campaigns are required. We report the discovery of \systemb\ ($=$\systemnb), a Saturn-mass transiting exoplanet orbiting a bright (T=10.0) Solar-type star (\teff=5730\,K).  \systemb\ was identified as a transiting exoplanet candidate from a single transit event of 1.3\% depth and 7.29\,h duration in \tess\ Sector 31 and a prior transit event from 2017 in NGTS data. Four further transit events were observed with \ngts\ photometry which revealed an orbital period of P=69.34\,days.  The transit events establish a radius for \systemb\ of $0.865 \pm 0.010\,\rm R_{\rm J}$, while radial velocity measurements give a mass of $0.386 \pm 0.025\,\rm M_{\rm J}$.  The equilibrium temperature of the planet is $414$\,K, making it much cooler than the majority of TESS planet discoveries. We also detect a transit signal in \ngts\ data not caused by \systemb, along with transit timing variations and evidence for a $\sim$150\,day signal in radial velocity measurements. It is likely that the system hosts additional planets, but further photometry and radial velocity campaigns will be needed to determine their parameters with confidence. \systemtb$\slash$\systemnb\ joins a small but growing population of cool giants that will provide crucial insights into giant planet composition and formation mechanisms.
\end{abstract}




\section{Introduction}

While thousands of transiting exoplanets are now known, the observed population is very strongly biased to hot, close-in planets with short orbital periods (typically $<$10\,d)\footnote{NASA Exoplanet Archive - accessed 2023-09-03}.  This is due to decreasing transit probability with wider orbital separation and the practical difficulty of detecting and confirming long-period transit signals.  The result of this bias is that most well studied giant planets have equilibrium temperatures greater than 1000\,K \citep[e.g.][]{2018AJ....156..103K}.  To study planetary atmospheres at temperatures closer to those in our own Solar System, we need to discover longer period transiting exoplanets.  Such planets will provide our best opportunities to characterise warm/cool planetary atmospheres using techniques such as transmission spectroscopy \citep[e.g.][]{2016Natur.529...59S} or phase curves \citep[e.g.][]{2021A&A...653A.173M}.  Atmospheric composition is expected to be sensitive to the formation location in the protoplanetary disc \citep[e.g.][]{2011ApJ...743L..16O,2020A&A...642A.229C} as well as the mechanisms driving planetary migration \citep[e.g.][]{2017MNRAS.469.4102M}.

Orbital obliquity (the misalignment between stellar rotation and orbital motion) can be measured via the Rossiter-McLaughlin effect \citep[e.g.][]{2010A&A...524A..25T} and is sensitive to the mechanism driving planetary migration \citep[e.g.][]{2012ApJ...757...18A}.
The handful of longer period transiting planets observed to date tend to be more aligned than the hot Jupiter population, potentially pointing to disc-driven migration rather than high-eccentricity migration \citep[e.g.][]{2021AJ....162...50W}.  

We cannot unravel the processes controlling the formation and evolution of exoplanets as a whole using only a sample limited to hot, close-in systems. More longer-period transiting planets, sampling a range of mass and radius, equilibrium temperature, eccentricity, age, and host star stellar type are required to increase the current small sample size when compared to the wealth of short-period planets. These planets are particularly valuable as their observed properties are much less affected by radiation from their host star, and their measured physical properties give insights into metal enrichment processes \citep[e.g.][]{2022AJ....163...61D,2022A&A...666A..46U} which are crucial to understand planetary formation history \citep{2016ApJ...831...64T,2020MNRAS.498..680G}.

Wide-field ground-based photometric surveys such as WASP \citep{2006PASP..118.1407P}, HAT-Net/HATSouth \citep{HATNet,HATSouth}, and KELT \citep{2007PASP..119..923P} seldom detect transiting exoplanet systems with periods longer than about 10\,days. Radial velocity surveys have been successful in spectroscopically characterising the orbit for many cool planets \citep[e.g.][]{2022ApJS..262...21F,2011A&A...535A..54S}. In particular, there have been instances of planets discovered by radial velocity measurements which have then been confirmed as a single-transit in \tess\ \citep[e.g.][]{2020MNRAS.496.4330D}. However, radial velocity surveys detect planets with a greater range of orbital inclinations and only a handful will be suitable for transit spectroscopy. 


The NASA Transiting Exoplanet Survey Satellite  \citep[\tess;][]{2015JATIS...1a4003R} mission is a wide-field photometric survey that is sensitive to exoplanet transits on bright stars across most of the sky.  \tess\ observes most stars for $\sim$27\,d during a single Sector, so planets with orbital periods between 13.5-27\,d might only exhibit a single transit event, and planets with periods longer than 27\,d cannot transit more than once per sector. Simulations show that hundreds of long-period planets around bright stars should be detectable as single transit events in \tess\ data, with orbital periods extending up to hundreds of days \citep{2018A&A...619A.175C}; similar results were found by \citet{2019AJ....157...84V} and later by \citet{2024MNRAS.529..715R}.  Single transits require extensive follow-up to determine the true orbital period.  High resolution spectroscopic follow-up observations are useful to constrain the mass and eccentricity, along with the orbital period.  However, there are a limited number of spectrographs on large telescopes capable of this follow-up, and these instruments are already over-subscribed monitoring multi-transit event TESS exoplanet candidates. Therefore, photometric follow-up plays an essential role in determining the true period for these single-transit event \tess\ candidates \citep{2021MNRAS.500.5088C}.  

In the extended \tess\ mission, most of the stars monitored in Year 1 or Year 2 were monitored for at least one additional Sector.  In some cases targets exhibit a second transit event in the extended mission, making the candidate a `duotransit' \citep[two monotransits separated by a wide data gap making the period ambiguous; ][]{2022A&A...664A.156O}.  In such cases a small number of possible periods are allowed, greatly reducing the amount of telescope time required for photometric follow-up.  However, for candidates with orbital periods longer than approximately 20\,days there are still many candidates that remain as single transit events after being observed in two \tess\ Sectors.  These candidates remain our best source of long period transiting planets from \tess, but also require the most demanding follow-up campaigns \citep[e.g.][]{2023A&A...674A..44G,2023MNRAS.523.3069O,2023MNRAS.523.3090T,2023A&A...674A..43U}.

In this paper we present one such successful campaign to follow-up the \tess\ single-transit candidate \systemt. A second transit was found in survey data taken three years earlier by the Next-Generation Transit Survey \citep[\ngts;][]{2018MNRAS.475.4476W}, and the orbital period was determined through subsequent long-term monitoring with NGTS. Consequently, we assign the names \systemtb\ and \systemnb\ to the planet, reflecting the complementary usage of the two facilities.
 
In Section \ref{sec:photometry} we set out the photometric observations of \system, both from \tess\ and NGTS.  In Section \ref{sec:spectroscopy} we describe the spectroscopic observations used to determine the mass of  \systemb.  In Section~\ref{sec:analysis} we determine parameters of the planet and host star from joint modelling of these datasets.  We present the orbital solution in Section \ref{sec:analysis:orbital} and investigate evidence for additional planets in Section \ref{sec:additional_planets}. Our results are discussed and summarised in Sections \ref{sec:discussion}  \& \ref{sec:conclusion}.


\section{Photometry}\label{sec:photometry}
\subsection{Single transit detection with \tess}\label{sec:tessphotometry}

\begin{figure}
    \centering
    \includegraphics[width=0.49\textwidth]{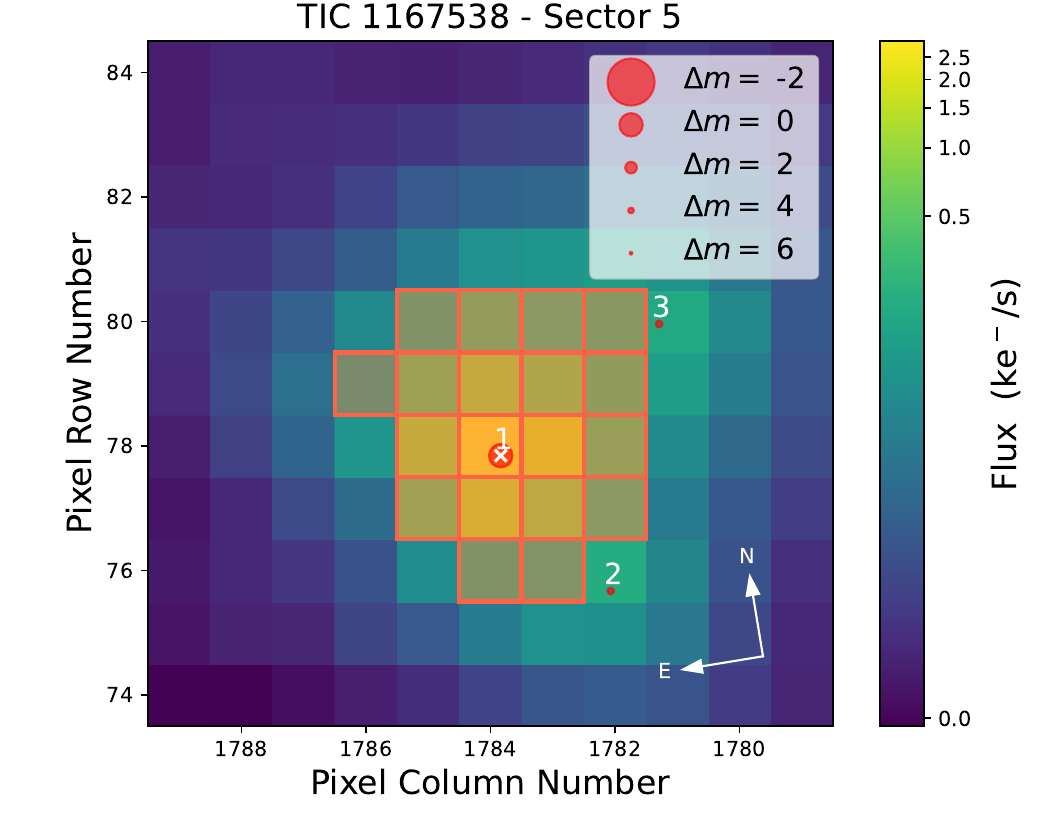}
    \caption{Target Pixel File \citep[TPF][]{2020A&A...635A.128A} from \tess\ sector 5 with \systemtic\ ($=$\systemt$\slash$\systemn) marked
    with a white cross. Other sources from \gaia\ DR3 are marked with red circles
    sized by scaled magnitudes relative to the target, ranked by distance. The
    aperture mask is indicated by the red outline.}
    \label{fig:TPF}
\end{figure}

\begin{figure}
    \centering
    \includegraphics[width=0.45\textwidth]{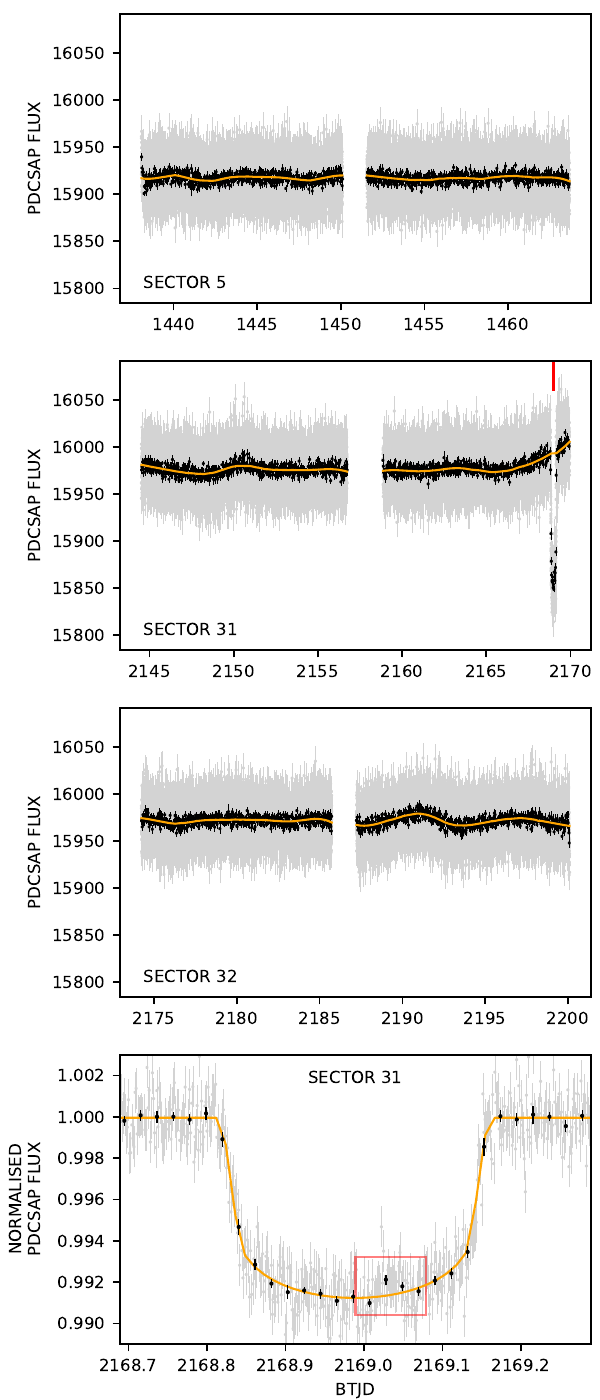}
    \caption{\tess\ SPOC 2-minute lightcurves plotted in raw cadence (grey points) and binned to 10 minutes (black points) for \systemtic$\slash$\systemt$\slash$\systemn\ from Sector 5 (top panel), Sector 31 (second panel) and Sector 32 (third panel).  The single transit event is marked in red in Sector 31.  A zoom-in of the normalised single-transit event in Sector 31 is shown in the lower panel, along with our best fitting transit model (orange line).  A possible spot crossing event can be seen during the transit (red box).}
    \label{fig:tess_lc}
\end{figure}

\systemtic\ is a bright (Tmag=10.01) G9V type dwarf star that was selected in the \tess\ candidate target list \citep[CTL;][]{2018AJ....156..102S} as a 2\,minute cadence target.  The star has parameters set out in Table \ref{tab:systemAparameters}. It was observed by \tess\ in Year 1 during Sector 5 (2018-11-15 to 2018-12-11) at 30-minite cadence, and again in Year 3 during Sectors 31 (2020-10-21 to 2020-11-19) and 32 (2020-11-19 to 2020-12-17) at 2-minute cadence; image data were reduced and analysed by the Science Processing Operations Center \citep[SPOC; ][]{2016SPIE.9913E..3EJ}.


Both the SPOC pipeline and the Quick Look Pipeline \citep[QLP; ][]{2020RNAAS...4..204H,2020RNAAS...4..206H} detected a single-transit event of 1.3 percent depth and 7.29 h duration in the TIC-1167538 light curve in Sector 31 centred at BJD=2459168.989. The TESS Science Office reviewed the data validation reports for the QLP detection and issued an alert for TOI-2447b on 6 January 2021 \citep{guerrero:TOIs2021ApJS}.

\tess\ photometry of \system\ is produced by SPOC and made publicly available on the Mikulski Archive for Space Telescopes (MAST)\footnote{https://mast.stsci.edu/}. We downloaded the 2-minute cadence SPOC HLSP data \citep{2020RNAAS...4..201C} from MAST which included Simple Aperture Photometry (SAP) extracted from the pipeline-derived photometric aperture \citep[Figure \ref{fig:TPF}; ][]{2010SPIE.7740E..23T,2017ksci.rept....6M} along with the Presearch Data Conditioning SAP \citep[PDCSAP;][]{Stumpe2012,Stumpe2014,2012PASP..124.1000S} light curve, which has been corrected for systematic trends shared by other stars on the detector (co-trending basis vectors) 
and is corrected for dilution.  We therefore use the SPOC PDCSAP lightcurve for the rest of this work.  \system\ shows significant photometric variability with a characteristic timescale of around 6\,days, likely due to star spots on the rotating stellar surface (Figure~\ref{fig:tess_lc}).  
To normalise the PDCSAP flux, we used an iterative algorithm that applied a Savitzky–Golay filter with a width of 2-days and rejected outliers until no further outliers were rejected from the previous iteration.
We excluded the event at BJD =2459168.989 and interpolate across the transit event to obtain a normalisation model for the whole dataset.  Our detrended and normalised lightcurve, along with a zoom in on the transit event, is presented in Figure~\,\ref{fig:tess_lc} and summarised in Table~\,\ref{tab:all_obs_summary}.  A prominent bump can be seen during the transit event of Sector 31 (see box in lower panel of Figure \ref{fig:tess_lc}).  This is likely due to a star spot on the facing hemisphere of \system\ that is occulted by \systemb.  Given the spot-induced photometric variability of \system\ (0.1-0.6\, ppt), it is not entirely unexpected to see such a spot-crossing event.  

We performed a preliminary fit of the transit event to check that \systemb\ was of planetary radius and to provide a model for the template matching set out in Sections~\ref{sec:ngts_archival_photometry} \& \ref{sec:ngtsphotometry}.  We use the method set out in \citet{2020MNRAS.491.1548G} to fit transit parameters assuming a 30-day orbital period and a circular orbit. These included the time of transit in the \tess\ data, the scaled orbital separation, the radius ratio, the impact parameter, a jitter parameter added in quadrature to the formal uncertainties from the SPOC data, and decorrelated limb-darkening coefficients for the power-2 law. We confirmed the transiting object was of a similar size to Saturn and obtained a good template model (Figure~\ref{fig:tess_lc}).

\begin{figure}
    \centering
    \includegraphics[width=0.5\textwidth]{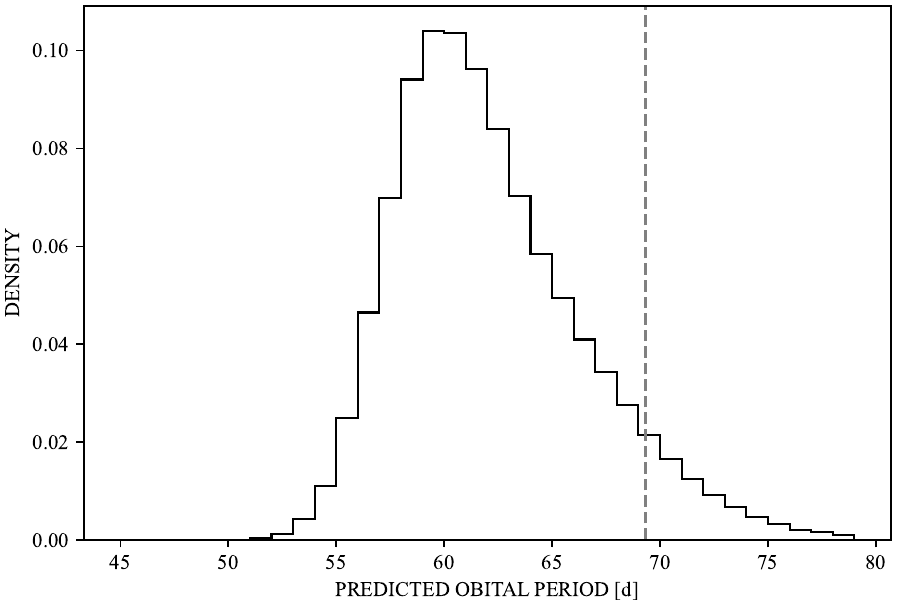}
    \caption{The posterior distribution of the predicted orbital period of \systemb\ (black solid) from a fit of the \tess\ single transit event. The true orbital period determined from NGTS photometery is shown (grey-dashed vertical line).  }
    \label{fig:SPOC_PERIOD}
\end{figure}

We were able to determine the posterior distribution of the orbital period from this preliminary modelling. For each valid trial step we calculated the transit duration using Eqn. 3 from \citet{2003ApJ...585.1038S}, generated a random value of the stellar density from a normal distribution centred on measured values from Table~\ref{tab:systemAparameters}. These were combined with impact parameter to estimate a posterior distribution for the true orbital period of \systemb\ using Equation 1 from \citet[][Figure~\ref{fig:SPOC_PERIOD}]{2015ApJ...815..127W}. From this, we estimated a probabilistic orbital period of $59^{+8}_{-4}\,\rm days$ (1-$\sigma)$ and that \systemb\ has a size similar to Saturn.



\subsection{\wasp\ archival photometry}\label{sec:wasp_archival_photometry}

We cross-matched \system\ with archival data from the Wide-Angle Search for Planets \citep[\wasp; ][]{2006PASP..118.1407P}.  \wasp operated two survey instruments: one at the South African Astronomical Observatory (SAAO), Sutherland, and another at the Observatorio del Roque de los Muchachos, La Palma. \system\ was observed for 2 consecutive observing seasons from 2006 to 2008 (1SWASPJ044359.41$-$315423.4) from the south station (13,918 observations in total). Data 
were detrended and aligned between  cameras and seasons using the SysRem algorithm \citep{Tamuz05} as implemented by \citet{2006MNRAS.373..799C}. These data do not have in-transit phase coverage for the solution of \systemb\ presented in this work, but they are useful for measuring the rotational period of the host star
(Section~\ref{section:rotation}). 
The signal-to-noise ratio on the timescale of an individual transit is not sufficient for a sensitive search for single transits of additional planets.

\subsection{\ngts\ archival photometry}\label{sec:ngts_archival_photometry}

The Next-Generation Transit Survey (NGTS) operates an array of twelve 20\,cm telescopes at the ESO Paranal Observatory in Chile.  Each \ngts\ telescope has been designed for high-precision photometry that matches \tess\ for all stars Tmag$>$12 (RMS=400\,ppm in 30\,min), and for stars with Tmag$>$9 by using multiple telescopes \citep[RMS=100\,ppm in 30\,mins; see][]{2020MNRAS.494.5872B}. Each of the 12 \ngts\ telescopes has a field-of-view of 8 square degrees, providing sufficient reference stars for even the brightest TESS candidates. The telescopes observe with a custom filter between 520-890\,nm and are specifically designed for precise photometry of exoplanet transits. The twelve independently mounted telescopes means that \ngts\ is one of the few ground-based facilities capable of monitoring multiple \tess\ single-transit objects simultaneously on a given night.
Our photometry is stable night-to-night and is capable of identifying exoplanet transits and stellar variability from night-to-night offsets \citep[e.g.][]{2023Natur.614..653A,2021MNRAS.504L..45B}. 

\begin{figure*}
    \centering
    \includegraphics[width=\textwidth]{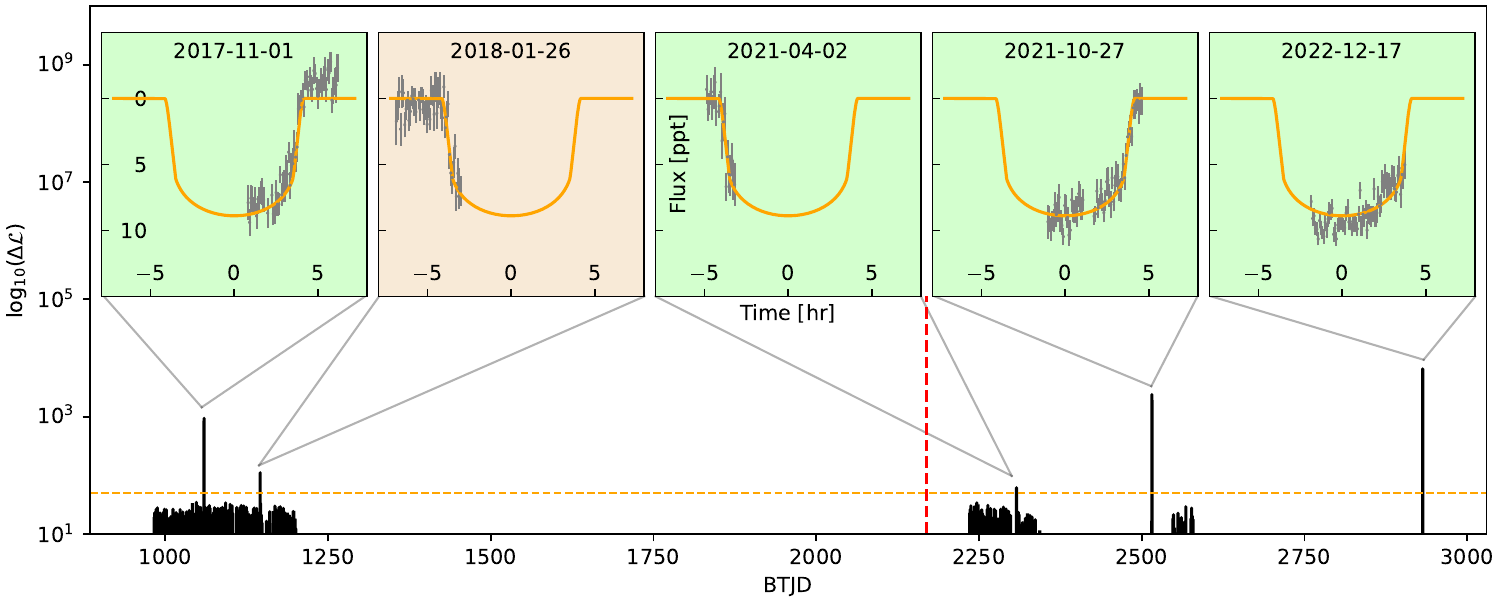}
    \caption{The $\Delta\mathcal{L}$ values (black histogram) for our template matching of the \tess\ Sector 31 transit event to \ngts\ data.  Inset plots show the five transit events that triggered above our threshold $\Delta\mathcal{L}$ value of 50 (horizontal and dashed orange line).  The inset plots have depth in ppt against time from mid-transit in hours, with the \tess\ template overlaid in orange.  The first two events were found in archival survey data, while the last three events were found by target monitoring.  The timing of the \tess\ single transit event (BTJD=2168.989) is marked with a red-dashed line. Green background inset panels are events from \systemb, while the red background inset panel is a candidate transit event from another planet.}
    \label{fig:ngts_detections}
\end{figure*}

       
\ngts\ obtained 203,854  data points  of \systemtic\ (=\systemt) across 175 nights from 2017-08-16 to 2018-03-23 as part of a survey of the Southern Hemisphere for transiting exoplanets. Data were reduced using standard aperture photometry routines and detrended for systematics as described in \citet{2018MNRAS.475.4476W}. Our photometric data is set out in Tables\,\ref{tab:all_obs_summary}\,\&\,\ref{tab:ngtsphot}. 
We used the template matching algorithm described by \citet{2020MNRAS.491.1548G} to search these observations for additional transits of \systemb. Through injection and recovery tests, we found a change in log-likelihood ($\Delta \mathcal{L}$) in excess of $\sim$50 corresponds to the significant detection of a transit event similar to the one seen in the \tess\ observations. 

We find a highly significant transit signal in the \ngts\ archival photometry (Figure \ref{fig:ngts_detections}) comprising almost the entire second half of the transit event centred at BTJD=1059.60676 (2017-11-01; $\Delta \mathcal{L}\sim 940$), which is more than three years earlier than the transit signal detected by \tess.
 The depth and egress shape are very well matched to the \tess\ event.
 
 We also find a second possible transit event centred at BTJD=1145.81594 (2018-01-26; $\Delta \mathcal{L}\sim 111$; second inset panel in Figure \ref{fig:ngts_detections}). This event does not have any data beyond the ingress, so we are unable to establish if it has the same depth as the \tess\ event.  We initially considered the possibility that  this event originated from the same planet. However this would imply that the orbital period was 86.2\, days, which can be excluded by the three sectors of \tess\ observations. Instead, we conclude that the 2018-01-26 event is most likely a real ingress of a second transiting planet.  The log-likelihood match to the transit shape of \systemb\ is high and the event is not correlated with any external observational parameters such as variations in background flux, PSF changes, changes in meteorological conditions, or telescope tracking/guiding issues.  The event does coincide with end-of-night/high-airmass; however, we do not see any other such events on previous or subsequent nights of observations.  Unfortunately, we do not find any further events in the \ngts\ or \tess\ photometry that confirm the second transiting planet or constrain its orbital period.  We discuss the possibility of an additional planet in the system in Section~\ref{sec:additional_planets}.

Based on the robust \ngts\ archival photometry transit of 2017-11-01 and the \tess\ transit event, we calculated that the true orbital period of planet b must be one of 16 aliases 
(1109.3848 / N)
with a lower limit of 
46.2238
days.



\subsection{NGTS Photometric Monitoring Campaign}\label{sec:ngtsphotometry}

\system\ was added to the \ngts\ monotransit follow-up program on 2021-01-14.  We observed the star with 10\,s exposures every possible night with a single \ngts\ telescope between 2021-01-14 and 2022-01-01, whenever the elevation of the field was above an airmass of 2.5.  Data were processed as described in Section \ref{sec:ngts_archival_photometry}, with the template matching algorithm 
automatically searching newly obtained \ngts\ photometric observations each night for transit events. 

On the night of 2021-04-02 an ingress event was detected that matched the depth and ingress shape of the original \tess\ single transit event ($\Delta \mathcal{L}~550$; Figure~\ref{fig:ngts_detections}). This event was uniquely consistent with alias N=16
(69.3357 days) 
of the 2017-11-01 \ngts\ archival event and thus we had determined the true orbital period of \systemb\ with photometry alone. 
Given the complementary use of the \tess\ and NGTS surveys to confirm the transit signal and measure the orbital period, we also assign the name \systemnb.

Interestingly, this third transit occurred eighteen minutes earlier than  expected based on the orbital period implied by first two transits, suggesting the possibility of transit timing variations in a multi-planet system.

We continued to monitor \system\ (\systemn) with \ngts\ to constrain the rotation of the host star from photometric modulation (Section~\ref{section:rotation}) and search for transit signals from additional planets. Exactly 5 cycles later on the night of 2021-10-27 \ngts\ observed a 
fourth
transit of \systemb\ ($\Delta \mathcal{L}>8000$) which covered in-transit and egress. 
This transit occurred 38\,min earlier than expected.
A fifth transit event of \systemb\ was observed 11 cycles later on the night of 2022-12-17 based on a targeted transit observation.  This transit contains entirely in-transit data ($\Delta \mathcal{L}>10,000$) and it occurred 26\,min earlier than expected. We discuss the detection of transit timing variations in Section~\ref{sec:ttvs}.


\begin{table}
\caption{Photometric colours, stellar atmospheric parameters, and physical properties of the host star \system$\slash$\systemn.}              
\label{tab:systemAparameters}      
\centering   
\begin{tabular}{l c c}          
\hline\hline                        
Parameter & value & Source\\
\hline 
TIC V8 ID & TIC-1167538 \\
TOI ID & TOI-2447 \\
\ngts & NGTS-29 \\
\gaia\ eDR3 Source ID & 4877544322252002048 & 1  \\
RA  &  $04^{\rm h} 43^{'}59.42^{"}$ & 1  \\
Dec  & $-39^{\circ} 54' 24^{"}$ & 1  \\

\\
pmRA [$\rm mas\, \rm yr^{-1}$] & $-0.294 \pm 0.011$  & 1  \\
pmDec [$\rm mas\, \rm yr^{-1}$] & $-31.651 \pm 	0.012$  & 1  \\
Parallax [$\rm mas$] & $6.6956 \pm  0.0115$  & 1  \\ 
Distance [pc] & $149.4 \pm 0.3$  & 1  \\ \\
Magnitudes\\
GAIA G & $10.4412 \pm 0.0028$  & 1  \\
GAIA BP & $10.763134 \pm 0.0028$  & 1  \\
GAIA RP & $9.956972 \pm 0.0037$  & 1  \\
TESS [T]  & $10.013 \pm 0.006$ & 2\\
APASS9 [B] & $11.199 \pm 0.087$ & 3\\
APASS9 [V] & $10.507 \pm 0.006$ & 3\\
2MASS [J]  & $9.485 \pm 0.024$ & 4 \\
2MASS [H]  & $9.165 \pm 0.021$ & 4 \\
2MASS [K$_{s}$]  & $9.12 \pm 0.026$ & 4 \\ \\

Spectroscopic parameters \\
$\rm T_{\rm eff}$ $\rm(K)$ & $5730 \pm 80$ & 5 \\
$\log g$ (dex)  & $4.3 \pm 0.1$ & 5 \\
$\xi_{\rm t}\, (\rm km\,s^{-1})$ & $1.08 \pm 0.18$ & 6 \\
$v_{\rm mac}\, (\rm km\,s^{-1})$ & $3.86 \pm 0.73$ & 6 \\
Vsin$i$ (km\,s$^{-1}$) & $3.5 \pm 0.6$ & 5 \\
$\rm [Fe/H]$ (dex) & $0.18 \pm 0.08$ & 5 \\ \\

Host parameters \\
$M_{\rm \star}$ [$M_{\odot}$] & $1.034 \pm 0.032$ & 5 \\
$R_{\rm \star}$ [$R_{\odot}$] & $1.006 \pm 0.009$ & 5 \\
Age [Gyr] & $2.1 \pm 1$ & 5 \\
Rotation period [d] & $\sim$13 & 5\\
\hline
\multicolumn{3}{l}{$^1$ \citet{2018A&A...616A...1G}, $^2$ \citet{2018AJ....156..102S},}\\
\multicolumn{3}{l}{$^3$ \citet{2015AAS...22533616H},  $^4$ \citet{2006AJ....131.1163S},}\\
\multicolumn{3}{l}{$^5$ this work, $^6$ value and uncertainties from \citet{2015PhDT........16D}.}
\end{tabular}
\end{table}

\begin{table}
\caption{Summary of photometric and radial velocity observations of \systemt$\slash$\systemn.}              
\label{tab:all_obs_summary}      
\centering   
\begin{tabular}{l c c}          
\hline\hline                        
Instrument & Number of observations & Time span \\
\hline
Photometry \\
\ngts\ (10 sec) & 203,854 & 2017-08-16 $\rightarrow$  2018-03-23   \\
\tess\ (2 min) & 8,669 & 2018-11-15 $\rightarrow$ 2020-12-16 \\
\ngts\ (10 sec) & 64,548 & 2021-01-15 $\rightarrow$ 2022-01-01 \\ \\

Spectroscopy \\
\coralie & 9 &  2021-01-21 $\rightarrow$ 2022-02-12\\
\harps & 34 & 2021-03-14 $\rightarrow$ 2022-02-14\\
\pfs & 4 & 2021-01-25 $\rightarrow$ 2022-01-16\\
\chiron & 78 & 2021-01-11 $\rightarrow$ 2021-12-19 \\
\feros & 10 & 2021-02-18 $\rightarrow$ 2021-11-23 \\
\hline
\end{tabular}
\end{table}

\begin{center}
\begin{table*}
    \centering
    \caption{Spectroscopic data for \systemt$\slash$\systemn. This table is available in its entirety online.}
    \label{tab:specdata}
    \begin{tabular}{l c c c c c c c c c c}  
    \hline
    \hline
    Instrument & Time (BJD & RV & RV error & FWHM & Bisector & Contrast & H-$\alpha$ &  Ca\,II H K &  Na\,D  \\
    & -2457000) & ($\rm m\,\rm s^{-1}$) & ($\rm m\,\rm s^{-1}$) & ($\rm m\,\rm s^{-1}$) & ($\rm m\,\rm s^{-1}$) &  & & &  \\
    \hline
    \coralie & 2235.71157 & 2992.13 & 28.22 & 8638.54 &-6.66 & 41.40786 & 0.196606 & -0.135634 & --\\
    \coralie & 2245.57602 & 2994.04 & 29.27 & 8582.32 &13.76 & 41.02233 & 0.181445 & 0.304748 & --\\
    \coralie & 2249.64778 & 2975.21 & 25.76 & 8613.98 &-0.56 & 40.74767 & 0.189266 & 0.099478 & --\\
    \coralie & 2472.85284 & 2990.56 & 19.28 & 8603.60 &-68.67 & 40.19499 & 0.201108 & 0.050821 & --\\
    \coralie & 2484.83206 & 2992.01 & 23.23 & 8605.20 &-25.00 & 40.19570 & 0.202674 & 0.174295 & --\\
    ... & ... & ...   & ...   & ...   & ...   & ...   & ...  & ...   & ... \\
\hline
    \end{tabular}
\end{table*}
\end{center}

\begin{center}
\begin{table}
    \centering
    \caption{\ngts\ photometric data for \systemt$\slash$\systemn. This table is available in its entirety online.}
    \label{tab:ngtsphot}
    \begin{tabular}{l c c c c }  
    \hline\hline
    Time (BJD   & Normalised flux & Flux uncertainty \\
    -2457000) & & \\
    \hline
981.89302 & 0.99536 & 0.00227 \\
981.89317 & 0.99775 & 0.00227 \\
981.89332 & 1.00385 & 0.00228 \\
981.89346 & 0.99793 & 0.00227 \\
981.89361 & 1.00359 & 0.00228 \\
981.89377 & 0.99790 & 0.00227 \\
981.89391 & 0.99601 & 0.00227 \\
981.89406 & 1.00066 & 0.00228 \\
981.89421 & 0.99692 & 0.00227 \\
    \multicolumn{1}{c}{...} & \multicolumn{1}{c}{...} & \multicolumn{1}{c}{...} \\
    \hline
    \end{tabular}
\end{table}
\end{center}

\section{Spectroscopy}\label{sec:spectroscopy}

To determine whether the transiting body is of planetary mass, we used spectroscopic radial velocity measurements as set out in the following sections. These observations are summarised in Table \ref{tab:specdata}.

\subsection{Vetting Spectroscopy with \coralie}\label{sec:coralie}

Many single transit events are in fact long period eclipsing binary stars \citep[e.g.][]{2020MNRAS.491.1548G,2020MNRAS.495.2713G,2020MNRAS.492.1761L,2021A&A...652A.127G,2022MNRAS.513.1785G}.  In order to rule out such systems, we use the \coralie\ instrument \citep{2001A&A...379..279Q} --- a fiber-fed \'{e}chelle spectrograph installed on the 1.2-m Leonard Euler telescope at the ESO La Silla Observatory in Chile.  With an exposure time of 40\,min in good conditions, \coralie\ can achieve a precision of 5-6\,\ms\ for bright solar-like stars \citep[e.g.][]{2010A&A...511A..45S}. 

We obtained a total of seven spectra with \coralie\ between 2021-01-21 and 2022-02-12, each with exposure times varying between 900 and 1200\,s.  The spectra were reduced using the standard \coralie\ reduction pipeline, and radial velocity measurements derived from standard cross-correlation techniques with a numerical G2 mask.  The data confirmed that the \system\ system had no large semi-amplitude that would indicate an eclipsing binary, so we scheduled the star for precision radial velocity follow-up (Section~\ref{sec:harps}).


\subsection{Radial Velocity monitoring with \harps}\label{sec:harps}

To determine the mass of \systemb, we used the \harps\ spectrograph \citep{2002Msngr.110....9P} on the 3.6\,m ESO telescope at La Silla Observatory in Chile.  In total 19 observations were made using \harps, with 8 between 2021-03-14 and 2021-10-30 by the WINE collaboration (PI:Brahm), and a further 11 observations between 2021-10-10 and 2022-03-15 by the Warm Jupiter program (108.22L8.001 PI: Ulmer-Moll).  Observations were taken using an exposure time of 1200\,s, which reached a mean signal-to-noise of 50 at 550\,nm.  \harps\ spectra were reduced using the standard \harps\ reduction pipeline with radial velocity measurements derived using the cross-correlation technique and a numerical G2 mask. 
The radial velocities 
are set out in Table~\ref{tab:specdata} and plotted in Figure~\ref{fig:TOI-2447_b_orbital}.

\subsection{Radial Velocity monitoring with \feros}

A total of 10 radial velocity measurements were obtained between 2021-02-18 and 2021-11-23 (PI Schlecker) using the \feros\ spectrograph in the context of the Warm gIaNts with tEss collaboration \citep[WINE, ][]{2019AJ....158...45B, 2021AJ....161..235H}. \feros\ is a stabilized high resolution spectrograph ($R=48,000$) installed on the MPG/ESO 2.2-m Telescope  \citep{1999Msngr..95....8K} in the ESO La Silla Observatory, in Chile. All \feros\ observations were performed with the simultaneous calibration technique with a ThAr lamp as the comparison source. The adopted exposure time was of 300\,s, which translated in spectra with signal-to-noise ration ranging from 60 to 80. The \feros\ data were reduced with the CERES \citep{2017PASP..129c4002B} pipeline which performs all steps required to obtain precision radial velocities starting from the raw images. The radial velocities were computed via cross-correlation with a G2-type binary mask. Bisector span measurements were also computed from the cross-correlation peak.  The CERES pipeline also performs an initial determination of the stellar atmospheric parameters from the continuum normalised spectra, obtaining in this case: \teff $= 5650.0 \pm 200$ K,	\logg $= 4.25 \pm 0.2$ dex, \feh$= 0.0 \pm 0.5$ dex, and \vsini $= 2.5 \pm 2.0$ km s$^{-1}$. These observations were consistent with the final orbital solution but not numerous or precise enough to measure the mass of \systemb\ and they were not used in our joint modelling. Nevertheless, the measurements are listed in full in Table~\ref{tab:specdata}.

\subsection{Radial Velocity monitoring with \chiron}

A total of 78 radial velocity measurements were obtained between 2021-01-11 and 2021-12-19 (PIs Brahm, Quinn, and Carleo) using the \chiron\ optical high-resolution ech\'{e}lle spectrograph \citep{2013PASP..125.1336T} on the 1.5-m telescope at the Cerro Tololo Inter-American Observatory. \chiron\ has a spectral resolution of R$\sim$79,000 across 415 to 880\,nm. All \chiron\ observations were reduced using the standard reduction pipeline based on the \textsc{reduce} package described in \citet{2002A&A...385.1095P}. Radial velocities were extracted using standard cross-correlation techniques using the observed spectrum as a template. The median uncertainty for \chiron\ radial velocity measurements is $\sim$13\,$\rm m\, \rm s^{-1}$. The radial velocities are set out in full in Table~\ref{tab:specdata} and plotted in Figure~\ref{fig:TOI-2447_b_orbital}.

\subsection{Radial Velocity monitoring with \pfs}

A total of four radial velocity observations and three exposures for a stellar template observation were obtained between 2021-01-25 to 2022-01-16 (PI Quinn) using the Planet Finder Spectrograph \citep[PFS;][]{crane2006, crane2008, crane2010} on the 6.5-meter Magellan II Clay telescope at Las Campanas Observatory in Chile. PFS is a slit-fed high-resolution echelle spectrograph operating at a spectral resolution of 130,000 with the $0.3\arcsec\times 2.5\arcsec$ slit, and its typical precision on bright RV standard stars is around 0.5--1.0~$\rm m \, \rm s^{-1}$. The observations of TOI-2447 were taken with a 3$\times$3 binning CCD readout mode to reduce read noise, and with an exposure time of 20 minutes per frame, the RV precision was around 1.1~$\rm m \, \rm s^{-1}$. The data were reduced and analysed for RV extraction using a customised pipeline \citep{butler1996}.  
These observations were consistent with the final orbital solution but not sufficiently numerous 
to measure the mass of \systemb\ and they were not used in our joint modelling. They are listed in full in Table~\ref{tab:specdata}.

\section{Analysis}\label{sec:analysis}

\subsection{Stellar atmospheric and physical parameters}\label{sec:analysis:host}

We used the \harps\ spectra to determine the parameters of the host star \systemt$\slash$\systemn, since \harps\ provided the highest signal-to-noise and sufficient resolution to measure line profiles. Each \harps\ spectrum was corrected into the laboratory reference frame
and co-added onto a common wavelength scale to create a high quality spectrum with signal-to-noise $\sim 153$. As described by \citet{2020ApJ...898L..11G}, a grid of pre-computed model spectra were synthesised with the software package \textsc{spectrum} \citep{1999ascl.soft10002G} using MARCS model atmospheres \citep{2008A&A...486..951G}, version 5 of the Gaia ESO survey (GES) atomic line list and solar abundances from \citet{2009ARA&A..47..481A}. Values of macroturbulence and microturbulence were calculated using equations 5.10 and 3.1 respectively from \citet{2015PhDT........16D}. Given these models, we used the H$\alpha$, NaI\,D, and MgI\,b lines to determine the stellar effective temperature, \teff, and surface gravity, \logg. Individual FeI and FeII lines provided a measurement of metallicity, \feh, and the rotational broadening projected into the line of sight, \vsini.

We used the method described by \citet{2020MNRAS.491.1548G} to determine the mass, radius, and age of \system.  This method uses Gaia magnitudes and parallax \citep{2018A&A...616A...1G} along with \teff\ and \feh\ from the spectroscopic analysis to determine the best-fitting stellar parameters with respect to MESA models \citep{2016ApJS..222....8D,2016ApJ...823..102C}. We found \system\ to be a main sequence G-type star that is $\sim2$\,Gyr old and has physical parameters consistent with the Sun. Our results are in good agreement with physical parameters listed in version 8 of the \tess\ input catalogue and those from the CERES pipeline used to reduce \feros\ data. The results of our analysis are presented in Table~\ref{tab:systemAparameters}. 


\subsection{The rotation and spin-orbit alignment of \system}\label{section:rotation}

\begin{figure}
    \centering
    \includegraphics[width=0.48\textwidth]{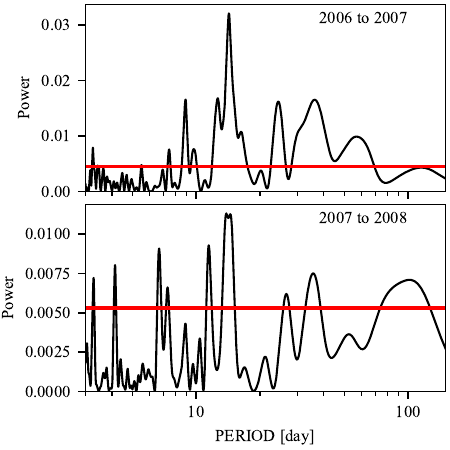}
    \caption{Lomb–Scargle power spectrum (black) of photometric observations for each season of \wasp\ data of \system. We also plot the power-spectrum level (red) corresponding to a 1\% false-alarm probability for the highest peak in each season of data.}
    \label{fig:WASP_LS}
\end{figure}

\begin{figure*}
    \centering
    \includegraphics[width = 1\textwidth]{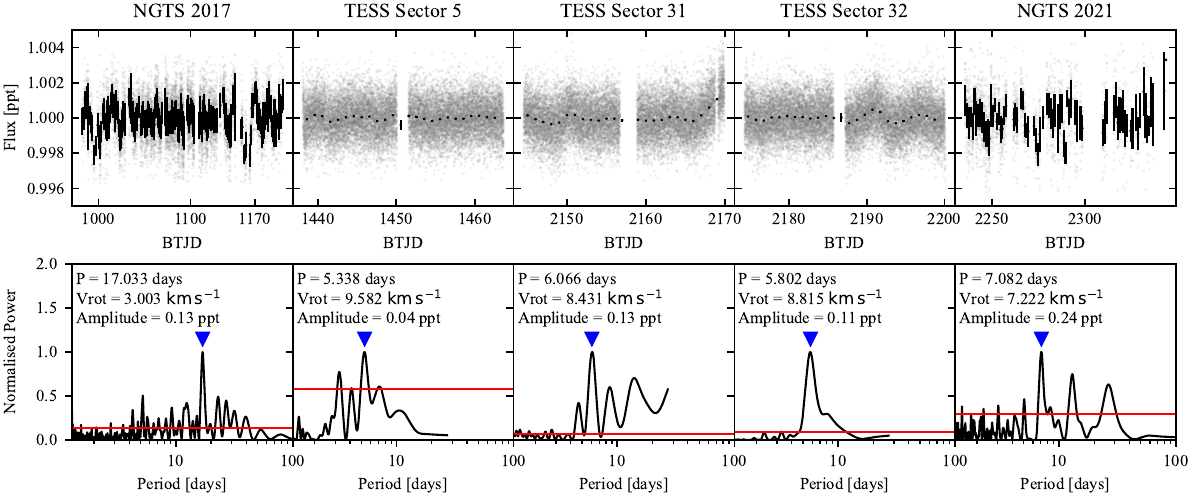}
    \caption{Lomb–Scargle analysis of photometric observations for independent \tess\ sectors (1 per column) using SPOC PDCSAP flux photometry and \ngts\ observations of \systemt$\slash$\systemn. Upper panels show the flux for each dataset with independent medians subtracted to avoid sector offsets. Lower panels show the Lomb–Scargle periodogram of the dataset (for the corresponding upper panel). In each case, we identify the peak of the Lomb-Scargle periodigram and note the corresponding 
    period and amplitude, along with the power-spectrum level (red) corresponding to a 1\% false-alarm probability for the highest peak. }
    \label{fig:SPOC_NGTS_LS}
\end{figure*}

In an attempt to determine the rotation period of the host star from rotational starspot modulation, we calculated Lomb-Scargle periodograms of the available photometric and spectroscopic data sets.

In Figure\,\ref{fig:WASP_LS} we show periodograms from two seasons of archival data from the WASP survey. Periodograms from both seasons show peaks at 14\,d, which correspond to an equatorial rotational velocity of $V_{ \rm rot} = 3.6\, \rm km\,\rm s^{-1}$. This is consistent with the \vsini\ measurement from our spectroscopic analysis (see Table\,\ref{tab:systemAparameters}) suggesting a $\sin i$ value close to unity and hence a planetary orbit aligned with the rotation of the star, as might be expected for a long period planet \citep[e.g.][]{2021AJ....162...50W}. The 14\,d peaks in Fig.\,\ref{fig:WASP_LS} are stronger than the 1\,\% false-alarm probability threshold, although multiple additional peaks above the threshold suggest the presence of significant red noise in the light curves. A second caveat is that a period of 14\,d is close to the first harmonic of  the Lunar month, which is associated with sky brightness variations that could potentially affect the WASP photometry \citep[e.g.][]{2014MNRAS.437.3133G} although we note there is no sign of the Lunar month in the periodograms of Fig.\,\ref{fig:WASP_LS}. 

In Figure\,\ref{fig:SPOC_NGTS_LS} we show Lomb-Scargle periodgorams for two seasons of photometric monitoring with NGTS and three sectors of \tess\ coverage. The 2017 NGTS data show a strong peak at 17\,d, which is consistent with the 14\,d period from WASP when considering differential stellar rotation and allowing for starspots emerging at different latitudes. The 2021 data have the strongest peak at 7\,d, which might be the first harmonic of a 14\,d period. We note that this second season of NGTS monitoring is significantly shorter than the 2017 season, and with more sparse coverage, reducing its sensitivity to longer periods.

The power spectra for the three \tess\ sectors in Fig.\,\ref{fig:SPOC_NGTS_LS} all show the strongest peaks at shorter periods of 5--6\,d. The duration of each \tess\ sector is only 27\,d, which is much shorter than the WASP and NGTS seasons, and is not well suited to the detection of a period as long as 14--17\,d. The \tess\ SPOC PDCSAP light curves have also been detrended for scattered light and instrumental artefacts, and it is possible power at 14-17\,d has been removed and the power at 5--6\,d represents red noise related to the residuals of the detrending. On the other hand, if the 5--6\,d peaks did represent the true rotation period of the star, this would imply an equatorial rotational velocity of around $9\, \rm km\,\rm s^{-1}$, which is faster than our measured \vsini\ and would imply a misaligned planetary orbit.

\begin{figure}
    \centering
    \includegraphics[width=0.5\textwidth]{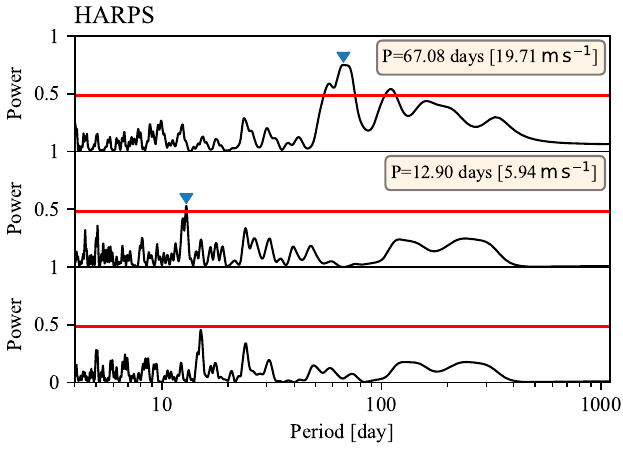}
    \includegraphics[width=0.5\textwidth]{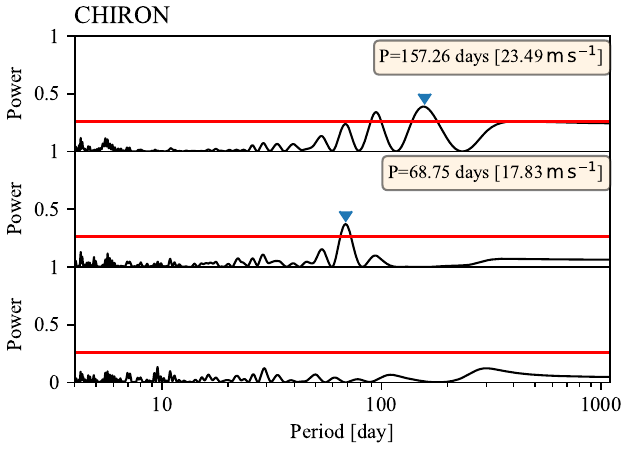}
    \caption{Upper panels -- the periodogram and prewhitening of prominent signals from a Lomb-Scargle analysis of \harps\ radial velocities. Signals above a false-alarm probability of 1\% were fit with sinusoids and removed until no peak existed above 1\%. Lower panels -- a similar analysis for the \chiron\ radial velocities.}
    \label{fig:RV_powers}
\end{figure}

We also calculated Lomb-Scargle power spectra for our
\harps\ and \chiron\ spectroscopic data, shown in Figure\,\ref{fig:RV_powers}. For each instrument we iteratively identify and the remove the strongest sinusoidal modulation until no signals with false-alarm probabilities below 1\,\% remain. For \harps, we first identify a strong peak at 68.79\,d with a false alarm probability below 0.001\%. This is the reflex motion caused by \systemb, and when this signal is removed from the data, we find one more peak at 12.9\,d (false alarm probability of $\sim$0.02\%). This is very close to the 14--17\,d peaks seen in WASP and NGTS photometry, and likely originates from stellar rotation. We also calculated power spectra for the \harps\ activity indicators listed in Table\,\ref{tab:specdata}, but none of these had peaks above the 1\% false alarm probability threshold. For \chiron\ observations, we first identify a signal at 157\,d
with a false alarm probability of $\sim$0.001\%, which may be the reflex motion of a second planet in the system
(see Section~\ref{sec:additional_planets}). We then see the signal at 69\,d caused by \systemb\ (false alarm probability of $\sim$0.003\%). We do not see the HARPS 12.9\,d signal with CHIRON, which is less sensitive than HARPS (but which can detect the 157\,d signal due to the longer baseline of observations).

Based on the detection of similar periods in WASP, NGTS and HARPS data, we conclude that the rotation period of the star is most likely the 13\,d signal detected with HARPS, which implies a rotational velocity consistent with our measured \vsini\ and hence an aligned orbit for  \systemb. The slightly longer periods of 14--17\,d detected with WASP and NGTS may reflect differential rotation of the star.

\subsection{Search for additional transiting planets}\label{sec:analysis:additional_planets}

The precision of \tess\ and \ngts\ photometery is such that we can visually detect transits from ice or gas giants such as \systemb. Template matching revealed a transit event in \ngts\ data which does not originate from \systemb\ (see Section \ref{sec:ngts_archival_photometry})
and thus began a search for additional transit signals in our data.  Visually, we do not find any additional transit events from similar sized objects above the $3$-$\sigma$ scatter of 3\,ppt observed in \tess\ observations or 6\,ppt for \ngts\ observations. 

As an additional check, we performed a Box-fitting Least Squares search \citep[BLS;][]{2002A&A...391..369K} of the detrended \tess\ PDCSAP flux, searching transit durations between 0.1 and 2 days and orbital periods between 2 and 300 days, allowing for single-transits. We found no significant peaks and therefore find no additional evidence of transiting planets in the \tess\ data. 
For a star of this brightness the \ngts\ observations have lower photometric precision than the \tess\ observations, although they do span a significantly longer baseline.  For the single \ngts\ camera the minimum recoverable depth at the noise limit is 0.6\,ppt, equating to a minimum recoverable planet radius above $\sim \rm 3.7 \, \rm R_{\oplus}$. We performed the same BLS search on \ngts\ observations and find no additional significant detections beyond the ingress event presented in Sect.\,\ref{sec:ngts_archival_photometry} and Fig.\,\ref{fig:ngts_detections}.


\subsection{Transit timing variations of \systemb}\label{sec:ttvs}

\begin{figure}
    \centering
    \includegraphics[width=0.5\textwidth]{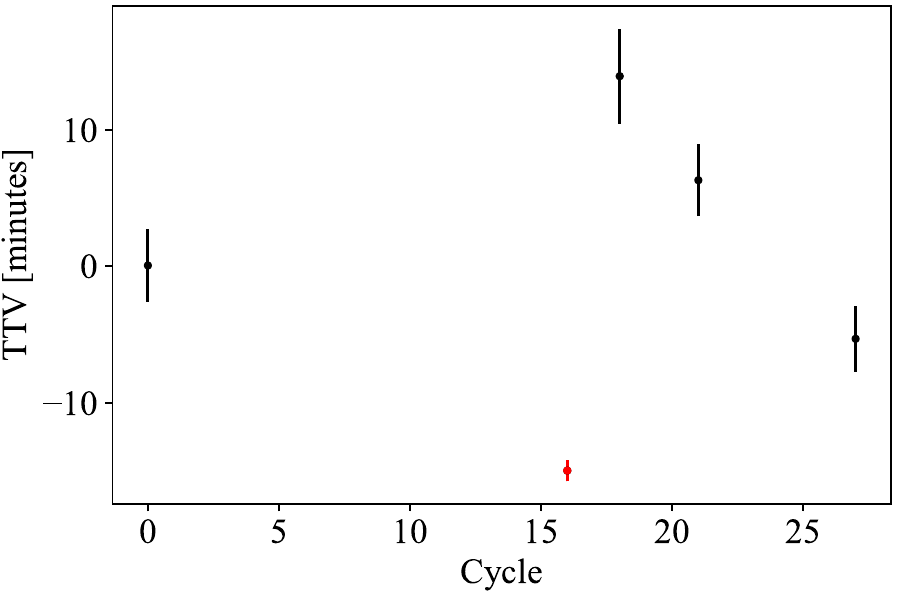}
    \caption{Modelling for the time of mid-transit of photometric transit events of \systemtb$\slash$\systemnb. The \tess\ single-transit event is shown in red with the \ngts\ events in black.}
    \label{fig:TOI-2447b_ttvs}
\end{figure}

We assessed the significance of the measured transit timing variations noted in Section \ref{sec:ngtsphotometry} by modelling the \tess\ and \ngts\ events together, with and without transit timing variations (summarised in Figure \ref{fig:appendix:ttv_justification}). We first fitted a common epoch and period, both fit simultaneously with scaled orbital separation, radius ratio, independent photometric zeropoints for each event and a jitter term for each instrument. We used Bayesian fitting described in Section~\ref{sec:tessphotometry} to find the best solution; however, we found this approach unable to accurately represent some ingress and egress events (left panel of Figure \ref{fig:appendix:ttv_justification}). Instead of fitting a common epoch and period, we then fitted the time of mid-transit for each event, and used the slope of a least squares fit to the transit times as a function of transit number to derive the precise orbital period of \systemb. Similarly to before, we fitted the scaled orbital separation, radius ratio, independent photometric zeropoints for each event with a jitter term for each instrument. We found this approach to better represent the transit events. To determine the statistical significance of this, we used the Bayesian information criterion (BIC) for each model, 
\begin{equation}
    BIC = \chi^2 + k \ln (n),
\end{equation}
where $k$ is the number of free parameters, $n$ is the number of points in the \tess\ and \ngts\ datasets, and $\chi^2$ is calculated using the model and data with jitter terms fixed at 0. Generally, the model with the lowest BIC is considered a better model; in this case the model with transit timing variations (495.6) compared to models with fitted ephemerides (958.1) despite having 3 more free parameters. The difference in these two numbers (462.5) is considerable and consistent with the transit timing variations being a much better approach. Thus for the remainder of this work, we chose to model individual transit times for each event (1 in \tess\ and 4 in \ngts) and report the orbital period of \systemb\ as the average orbital period between transits. 
A summary of the transit timing variations for the final orbital solution is shown in Figure~\ref{fig:TOI-2447b_ttvs}  and Table~\,\ref{tab:derived_parameters}.

\section{Orbital solution of \systemtb$\slash$\systemnb}\label{sec:analysis:orbital}

\begin{figure}
    \centering
    \includegraphics[width=0.46\textwidth]{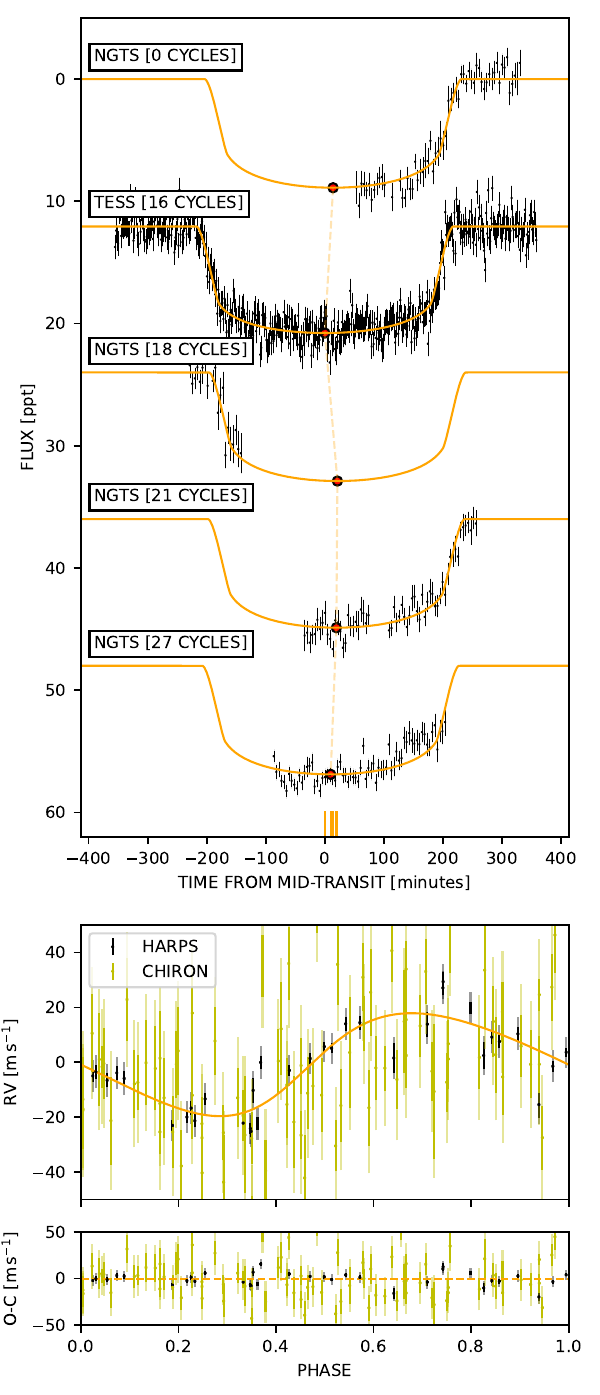}
    \caption{Orbital solution for \systemtb$\slash$\systemnb. The upper panel shows transit photomety from  \tess\ and \ngts\ with best fitting models (orange). Fitted times of transit centre for each event are marked with a red dot (joined with an orange-dashed line) and a corresponding orange tick on the bottom axis. The lower panels show radial velocity measurements with the best-fitting model (orange) with residuals for each radial velocity dataset. Semi-transparent error bars represent uncertainties with jitter values added in quadrature.}
    \label{fig:TOI-2447_b_orbital}
\end{figure}

We modelled all photometric and radial velocity datasets simultaneously following the methodology set out by \citet{2020MNRAS.491.1548G}. We decided to select the \tess\ transit as our reference epoch, and label each transit as its orbital cycle from the first \ngts\ event on 2017-11-01 -- $T_{C 0}$, $T_{C 16}$, $T_{C 18}$, $T_{C21}$, $T_{C27}$ (shown visually in Figure~\ref{fig:TOI-2447_b_orbital}). The trial orbital period, $P$, is determined from the gradient of a linear fit of these epochs with orbital cycle. Our fitted parameters also included $R_\star / a$, $k$, $b$, independent values of the photometric zero-point, $zp$,  and decorrolated limb-darking parameters $h_1$ and $h_2$  for each photometric dataset which represent a star limb-darkened by the power-2 law, the semi-amplitude of the planet, $K$, and the systematic radial velocity of the primary star, $V_0$. We avoided fitting the 
eccentricity 
($e$) and the argument of the periastron ($\omega$), which are strongly correlated, and instead used $f_{c} = \sqrt{e} \cos \omega$ and  $f_{s} = \sqrt{e} \sin \omega$ since these are less correlated and have more uniform prior probability distributions. Radial velocity errors are occasionally underestimated in-part due to stellar activity, pulsations, and granulation which can introduce noise in to the radial velocity measurements \citep{2006ApJ...642..505F}. To mitigate this, we include independent jitter terms, $J$, for each radial velocity data set which are added in quadrature with radial velocity errors. We fit a similar term for each photometric data set, $\sigma$, which was also added in quadrature to photometric uncertainties. We fit $h_1$ \& $h_2$ with Gaussian priors centred on tabulated values based on \system's stellar atmospheric parameters. The subtle differences between \tess\ and \ngts\ transmission filters are such that we fitted independent values of $h_1$ and $h_2$ for each photometric dataset. As part of the SPOC pipeline, the PDCSAP lightcurve has been corrected assuming a contamination ratio of 0.0205\% (calculated from TIC V8). The \ngts\ aperture is sufficiently isolated that no dilution correction is applied.

As described by \citet{2020MNRAS.491.1548G}, we explored the parameter space with a Bayesian sampler \textsc{emcee} \citep{2013PASP..125..306F} and drew 100,000 steps from 52 walkers (twice the number of fitted parameters) and discarded the first 50,000 steps as the burn-in phase. After visually confirming each chain had converged, we selected the trial step with the highest log-likelihood as our measurement for each fitted parameter. Asymmetric uncertainties were calculated from the difference between each measured parameter and the $16^{\rm th}$ and $84^{\rm th}$ percentiles of their cumulative posterior probability distributions. The best fitting solution is shown in Figure~\ref{fig:TOI-2447_b_orbital} with parameter values detailed in Table~\ref{tab:fitted_parameters}.

For each valid trial step in our joint fit, we calculate derived parameters that are of interest. We first calculate the transit duration using Eqn. 3 from \citet{2003ApJ...585.1038S}. To calculate the mass and radius of the planet, we draw random values of $M_\star$ and $R_\star$ from a normal distribution centred on measured values from Table \ref{tab:systemAparameters} with width equal to their respective uncertainties. These were combined with $P$, $e$, and $K_\star$ to make a closed-form solution of the cubic polynomial required to solve the mass function 
%
%
for the mass of the planet, $M_p$. The mass ratio, $q = M_p / M_\star$, can then be used with $R_\star / a$, $f_s$, and $f_c$ to calculate the surface gravity of the planet using Eqn. 4 from \citep{2007MNRAS.379L..11S}.
The radius of the star, $R_\star$ and $k$ were combined to calculate the radius of the planet $R_p$. Furthermore, the semi-major axis, $a$, was calculated by combining $R_\star$ with $R_\star/a$. Finally, $R_\star/a$ were combined with random values of \teff\ from a normal distribution of centre and width 5730\,K and 80\,K respectively (from Table \ref{tab:systemAparameters}) to calculate the equilibrium temperature of \systemb\
%
%
assuming a Jupiter-like bond albedo of 0.34 \citep[e.g.][]{2022AJ....163...61D}. The derived values from our joint analysis are presented in  Table \ref{tab:derived_parameters}.

\begin{figure}
    \centering
    \includegraphics{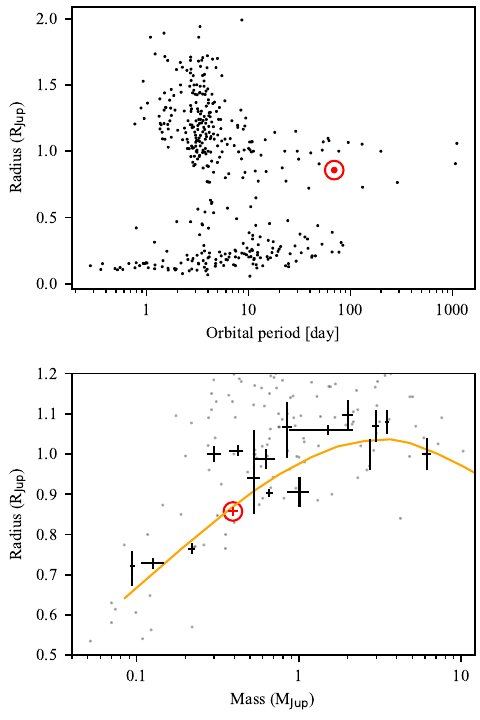}
    \caption{Upper panel: Radius-period diagram of well-characterised transiting exoplanets (mass to better than 50\% precision and radius to better than 20\%; exoplanetarchive.ipac.caltech.edu, accessed 2022-10-07). \systemtb$\slash$\systemnb\ is indicated in red. 
    Lower panel: Mass-radius diagram for giant exoplanets from the same sample. Planets with orbital periods longer than 30\, days are plotted in black, those with periods shorter than 30\,days in grey. \systemb\ is indicated in red with an outer circle. 
    The orange line shows a theoretical mass-radius relation for a cold hydrogen/helium exoplanet \citep{2007ApJ...669.1279S}.}
    \label{fig:TOI-2447_b_MR}
\end{figure}

\begin{table}
\caption{Orbital solution for \systemtb$\slash$\systemnb\ using a single-planet model. Asymmetric errors are reported in brackets and correspond to the difference between the median and the 16$^{\rm th}$ (lower value) and 84$^{\rm th}$ (upper value) percentile.}              
\label{tab:fitted_parameters}      
\centering   
\begin{tabular}{l c}          
\hline\hline                        
Parameter & value\\
\hline 
Fitted parameters \\

$\rm T_{\rm 0}$ [BTJD] & $1059.60965_{(210)}^{(198)}$  \\
$\rm T_{\rm 16}$ [BTJD] & $2168.98798_{(61)}^{(63)}$  \\
$\rm T_{\rm 18}$ [BTJD] & $2307.67868_{(271)}^{(278)}$  \\
$\rm T_{\rm 21}$ [BTJD] & $2515.68780_{(184)}^{(182)}$  \\
$\rm T_{\rm 27}$ [BTJD] & $2931.70114_{(217)}^{(221)}$  \\
$R_{\star} / a$  & $0.01285_{(104)}^{(109)}$  \\
$R_{p} / R_{\star}$ & $0.08704_{(82)}^{(97)}$  \\
$b$ & $0.277_{(138)}^{(104)}$  \\
$zp_{\rm 0}$ & $1.00098_{(13)}^{(13)}$  \\
$zp_{\rm 16}$ & $0.99994_{(7)}^{(7)}$  \\
$zp_{\rm 18}$ & $1.00009_{(33)}^{(33)}$  \\
$zp_{\rm 21}$ & $1.00017_{(16)}^{(15)}$  \\
$zp_{\rm 27}$ & $0.99958_{(13)}^{(13)}$  \\
$\rm h_{\rm 1, \ngts}$ & $0.758_{(3)}^{(3)}$  \\
$\rm h_{\rm 2, \ngts}$ & $0.777_{(41)}^{(40)}$  \\
$\rm h_{\rm 1, \tess}$ & $0.785_{(3)}^{(3)}$  \\
$\rm h_{\rm 2, \tess}$ & $0.765_{(41)}^{(43)}$  \\
$\sigma_{\rm TESS}$ & $0.00034_{(14)}^{(12)}$  \\
$\sigma_{\rm NGTS}$ & $0.00039_{(8)}^{(8)}$  \\
$V_0$ \harps\ [m\,s$^{-1}$] & $3030.53_{(107)}^{(103)}$  \\
$V_0$ \chiron\ [m\,s$^{-1}$] & $1560.33_{(209)}^{(214)}$  \\
$K_\star$ [m\,s$^{-1}$] & $18.8_{(14)}^{(14)}$  \\
$f_{\rm s}$ & $-0.403_{(89)}^{(136)}$  \\
$f_{\rm c}$ & $-0.110_{(73)}^{(60)}$  \\
$J_{\rm HARPS}$ [m\,s$^{-1}$] & $5.40_{(0.82)}^{(0.98)}$  \\
$J_{\rm CHIRON}$ [m\,s$^{-1}$] & $19.48_{(1.85)}^{(2.05)}$  \\
\hline
\end{tabular}
\end{table}

\begin{table}
\caption{Derived properties of \systemtb$\slash$\systemnb. Asymmetric errors are reported in brackets and correspond to the difference between the median and the 16$^{th}$ (lower value) and 84$^{th}$ (upper value) percentile.}              
\label{tab:derived_parameters}      
\centering   
\begin{tabular}{l c}          
\hline\hline                        
Parameter & value\\
\hline 
TTV T$_{\rm 0}$ [min] & $1.26_{(3.28)}^{(3.09)}$ \\
TTV T$_{\rm 16}$ [min] & $-16.13_{(0.96)}^{(0.98)}$ \\
TTV T$_{\rm 18}$ [min] & $10.43_{(4.23)}^{(4.33)}$ \\
TTV T$_{\rm 21}$ [min] & $8.24_{(2.87)}^{(2.85)}$ \\
TTV T$_{\rm 27}$ [min] & $-3.80_{(3.39)}^{(3.44)}$ \\
Period [day] & $69.33684_{(11)}^{(9)}$  \\
$M_{\rm p}$ [\mjup] & $0.393_{(27)}^{(31)}$  \\
$R_{\rm p}$ [\rjup] & $0.857_{(11)}^{(12)}$  \\
duration [hr] & $7.16_{(14)}^{(86)}$  \\
$g_{\rm p}$\,$\rm [m \, \rm s^{-2}]$ & $15.5_{(3.8)}^{(1.5)}$  \\
$e$ & $0.17_{(10)}^{(2)}$  \\
$\omega$\,[rad] & $-1.83_{(37)}^{(10)}$  \\
$\rm T_{\rm eq}$ [K]$^{1}$ & $414_{(7)}^{(28)}$  \\
$a$ [au] & $0.347_{(26)}^{(29)}$  \\
\hline
\multicolumn{2}{l}{$^1$ Assuming a bond albedo of 0.34}\\
\end{tabular}
\end{table}

\section{Evidence for additional planets}\label{sec:additional_planets}

It is most likely that \systemtb$\slash$\systemnb\ is not the only planet in this system. 

Archival \ngts\ data shows a partial transit event of unknown depth that cannot be attributed to \systemb\ (Sect.\,\ref{sec:ngts_archival_photometry}; Fig.\,\ref{fig:ngts_detections}), suggesting there is a second transiting planet in the system. This event was not found to repeat in the \tess\ or \ngts\ data. 
From Sectors 31 and 32 of \tess\ data, we can determine the minimum period of the additional planet to be $\geq$56\, days (shorter periods from the gaps in these data are excluded by Sector 5 data). This lower limit is close to the 69\,d period of planet b, and stability considerations suggest that the period of the additional transiting planet must be substantially longer. \ngts\ data exclude most periods between 60--80\,d and 100--120\,d.

We also have evidence of an object dynamically interacting with \systemb, causing 
transit timing variations (see Section \ref{sec:ttvs}). 
We do not have enough data to model the transit timing variations or an associated super-period; however, we are continuing to monitor \system\ with \ngts\ to measure more transit timing variations and search for additional transits that are not from \systemb. 

Because spectroscopic observations are sensitive to both transiting and non-transiting planets, we examined carefully the reflex motion seen in \chiron\ and \harps\ observations (Figure~\ref{fig:RV_powers}). For \harps, we find prominent signals at 68.79 days caused by the reflex motion of \systemb, and 12.9 days suspected to be caused by stellar rotation (see Sect.\,\ref{section:rotation}).  For \chiron\ observations, we first identify a signal at 157 days followed by a signal at 69 days caused by \systemb. The 157-day signal could be 
due to a wider separation planet that also accounts for the additional \ngts\ transit and/or the detected transit timing variations.



We continue to monitor the \system\ with photometry and spectroscopy in order to better characterise these signals and determine the properties of additional planets in the system.

\section{Discussion}\label{sec:discussion}

Our joint fit to photometric observations from \tess\ and NGTS, and radial velocity observations from \harps\ and \chiron\ show \systemtb$\slash$\systemnb\ has a mass and radius between Saturn and Jupiter, an orbital separation of $0.35$\,au and a low orbital eccentricity 
($\leq$0.19; our full set of parameters is presented in Tables\,\ref{tab:fitted_parameters}\,\&\,\ref{tab:derived_parameters}). 

We place \systemb\ in the context of the wider exoplanet population in Figure~\ref{fig:TOI-2447_b_MR}. The planet is consistent with the theoretical mass-radius relationship for a cold hydrogen/helium exoplanet \citep{2007ApJ...669.1279S}. \systemb\ has a surface gravity of $\rm 15 \,m\,s^{-2}$ and an equilibrium temperature of $414$\,K assuming a Jupiter-like Bond albedo. This is similar to the equilibrium temperature of NGTS-11\,b \citep{2020ApJ...898L..11G}.

\systemb\ joins a growing set of gas giants with equilibrium temperatures below 500\,K. Photometric transits provide useful insights into the bulk properties of exoplanets but spectroscopic transits provide atmospheric and obliquity constraints useful for testing models of planet formation. 
Based on the stellar and planetary parameters set out in Tables\,\ref{tab:systemAparameters}\,\&\,\ref{tab:fitted_parameters} and an aligned spin-orbit axis, we would expect the amplitude for the Rossiter-McLaughlin signal to be $\sim26\, \rm m \, \rm s^{-1}$ \citep[Eqn. 40 from][]{2010exop.book...55W}. This is well within the capabilities of many ground-based spectrographs for a Tmag=10 host star.  Thus we believe it will be possible to measure the spin-orbit alignment 
of\systemb, which we find 
is most likely to be aligned, based on our spectroscopic \vsini\ and most likely rotation period of 13\,d
(see Sect.\,\ref{section:rotation}).

Spectroscopic transits of warm Jupiters can also reveal how their atmospheres differ from their hot counterparts.  Assuming a Saturn-like mean molecular weight for \systemb\ (2.07\,g mole$^{-1}$)\footnote{https://nssdc.gsfc.nasa.gov}, we estimate an atmospheric scale height, $H_p$, of $121\pm18$ km for \systemb. 
The recently commissioned \jwst\ provides a unique opportunity to obtain near-infrared transit spectroscopy not achievable from the ground.  We used Eqn.\,1 from \citet{2018PASP..130k4401K} to calculate the transmission spectroscopy metric of 45 using values from Tables \ref{tab:systemAparameters} \& \ref{tab:derived_parameters}, which is high for a cool Jupiter. 
\jwst\ observations will inform our understanding the formation of cool giant planets, particularly those which have undergone dynamical migration,  constraining evolutionary models that are routinely used to describe exoplanets as a whole.

We find strong evidence of additional planets beyond \systemb\ including TTVs, an additional partial transit from NGTS, and a radial velocity signal at 157\,d in \chiron\ data. If the orbital period suggested by \chiron\ data is correct, this second planet resides reasonably close to the 2:1 mean motion resonance which is possible with convergent disk migration due to disk torques \citep[e.g][]{2002ApJ...567..596L}.

\section{Conclusion}\label{sec:conclusion}

A single transit event was detected around the Solar-like star \systemtic\ in \tess\ Sector 31, which subsequently led to its designation as a \tess\ object of interest (\systemt). 
Archival observations from \ngts\ revealed a second transit event, observed three years earlier, and subsequent NGTS monitoring detected a further three transits. These \ngts\ observations determined an orbital period of 69.34\,d for \systemtb$\slash$\systemnb, along with significant transit timing variations.  Radial velocity measurements were obtained with \coralie, \harps, \feros, \chiron, and \pfs, confirming a planetary mass. Joint modelling revealed the planet to have a mass of $0.393 \pm 0.031\,\rm M_{\rm J}$ and a radius of $0.857 \pm 0.012\,\rm R_{\rm J}$ with an equilibrium temperature of $414 \pm 28$\,K (assuming a Jupiter-like albedo). 
\systemb\ joins a growing population of warm Jupiters that will provide crucial insights into giant planet formation mechanisms through Rossiter-McLaughlin measurements and \jwst\ observations.

We find strong evidence for additional planets in this system. The 2018-01-26 ingress event from archival \ngts\ data does not originate from \systemb\ and likely originates from another transiting planet. We also find transit timing variations for \systemb\ suggesting there is some dynamical planet-planet interaction. There is also evidence in the \chiron\ radial-velocity data of an outer planet with an orbital period of around $150$\, days.
The signal is not detected with \harps, which is more sensitive but covers a shorter baseline. 
Further observations of this system are required in order to better characterise these signals and confirm the presence of additional planets.

\section*{Acknowledgements}

The \ngts\ facility is operated by the consortium institutes with support from the UK Science and Technology Facilities Council (STFC) under projects ST/M001962/1, ST/S002642/1 and ST/W003163/1. 
This work has been carried out within the framework of the National Centre of Competence in Research PlanetS supported by the Swiss National Science Foundation under grants 51NF40\_182901 and 51NF40\_205606. The authors acknowledge the financial support of the SNSF. We acknowledge the use of public TESS data from pipelines at the TESS Science Office and at the TESS Science Processing Operations Centre.
This paper includes data collected with the TESS mission, obtained from the MAST data archive at the Space Telescope Science Institute (STScI). Funding for the TESS mission is provided by the NASA Explorer Program. STScI is operated by the Association of Universities for Research in Astronomy, Inc., under NASA contract NAS 5–26555.
Based on observations made with ESO Telescopes at the La Silla Paranal Observatory under programme IDs $0104.C-0413$ (PI RB), $0104.C-0588$ (PI FB), Opticon:2019A/037 (PI DB), and CNTAC: $0104.A-9012$ (PI JIV).
The contributions at the University of Warwick by SG, DB, PJW, RGW and DA have been supported by STFC through consolidated grants ST/P000495/1, ST/T000406/1 and ST/X001121/1.
Contributions at the University of Geneva by SU, ML, FB, MB, AFK and SU were carried out within the framework of the National Centre for Competence in Research ``PlanetS'' supported by the Swiss National Science Foundation (SNSF) under grants 51NF40 182901 and 51NF40 205606. ML acknowledges support of the Swiss National Science Foundation under grant number PCEFP2 194576.
This research has made use of NASA's Astrophysics Data System Bibliographic Services and the SIMBAD database, operated at CDS, Strasbourg, France. This research made use of Astropy,\footnote{http://www.astropy.org} a community-developed core Python package for Astronomy \citep{2018AJ....156..123A}.
JSJ greatfully acknowledges support by FONDECYT grant 1201371 and from the ANID BASAL project FB210003.
This research has made use of the NASA Exoplanet Archive, which is operated by the California Institute of Technology, under contract with the National Aeronautics and Space Administration under the Exoplanet Exploration Program.
The contributions at the Mullard Space Science Laboratory by E.M.B. have been supported by STFC through the consolidated grant ST/W001136/1.
P.D. acknowledges support from a National Science Foundation (NSF) Astronomy and Astrophysics Postdoctoral Fellowship under award AST-1903811 and a 51 Pegasi b Postdoctoral Fellowship from the Heising-Simons Foundation.
The results reported herein benefited from collaborations and/or information exchange within the program “Alien Earths” (supported by the National Aeronautics and Space Administration under agreement No. 80NSSC21K0593) for NASA’s Nexus for Exoplanet System Science (NExSS) research coordination network sponsored by NASA’s Science Mission Directorate.
R.B. acknowledges support from FONDECYT Project 11200751 and from project IC120009 “Millennium Institute of Astrophysics (MAS)” of the Millenium Science Initiative.
A.J. acknowledges support from FONDECYT Project 1210718 and from project IC120009 “Millennium Institute of Astrophysics (MAS)” of the Millenium Science Initiative.
J.H. supported by the Swiss National Science Foundation (SNSF) through the Ambizione grant PZ00P2 180098.
EG gratefully acknowledges support from the UK Science and Technology Facilities Council (STFC; project reference ST/W001047/1).
T.T. acknowledges support by the DFG Research Unit FOR 2544 "Blue Planets around Red Stars" project No. KU 3625/2-1. T.T. further acknowledges support by the BNSF program "VIHREN-2021" project No. KP-06-DV/5.
D. D. acknowledges support from the NASA Exoplanet Research Program grant 18-2XRP18 2-0136, and from the TESS Guest Investigator Program grants 80NSSC22K0185 and 80NSSC23K0769.
This work made use of \texttt{tpfplotter} by J. Lillo-Box (publicly available in www.github.com/jlillo/tpfplotter), which also made use of the python packages \texttt{astropy}, \texttt{lightkurve}, \texttt{matplotlib} and \texttt{numpy}.
We thank the reviewer for their helpful comments that have improved this manuscript.

\section*{Data Availability}

\tess\ SPOC data is publically available to download from Mikulski Archive for Space Telescopes. Reduced \harps\ spectra, derived measurements of radial velocities, and the full photometric dataset from \ngts\ will be available from the VizieR archive server hosted by the Universit\'{e} de Strasbourg.\footnote{cdsarc.u-strasbg.fr}



\bibliographystyle{mnras}
\bibliography{paper} 




\appendix

\section{Affiliations}
$^{1}$ Department of Physics, University of Warwick, Gibbet Hill Road, Coventry CV4 7AL, UK\\
$^{2}$ Centre for Exoplanets and Habitability, University of Warwick, Gibbet Hill Road, Coventry CV4 7AL, UK\\
$^{3}$ Observatoire de Gen{\`e}ve, Universit{\'e} de Gen{\`e}ve, Chemin Pegasi 51, 1290 Versoix, Switzerland\\
$^{4}$ Facultad de Ingeniera y Ciencias, Universidad Adolfo Ib\'{a}\~{n}ez, Av. Diagonal las Torres 2640, Pe\~{n}alol\'{e}n, Santiago, Chile\\
$^{5}$ Millennium Institute for Astrophysics, Chile\\
$^{6}$ Data Observatory Foundation, Chile\\
$^{7}$ Departamento de Astronom\'ia, Universidad de Chile, Casilla 36-D, Santiago, Chile\\
$^{8}$ Centro de Astrof\'isica y Tecnolog\'ias Afines (CATA), Casilla 36-D, Santiago, Chile\\
$^{9}$ School of Physics and Astronomy, University of Leicester, Leicester LE1 7RH, UK\\
$^{10}$ Carnegie Earth and Planets Laboratory, 5241 Broad Branch Road NW, Washington, DC 20015, USA\\
$^{11}$ Mullard Space Science Laboratory, University College London, Holmbury St Mary, Dorking, Surrey, RH5 6NT, UK\\
$^{12}$ Center for Astrophysics | Harvard \\\
$^{13}$ The Observatories of the Carnegie Institution for Science, 813 Santa Barbara St., Pasadena, CA 91101, USA\\
$^{14}$ Instituto de Astrof\'{i}sica de Canarias (IAC), 38205 La Laguna, Tenerife, Spain\\
$^{15}$ Department of Astronomy and Astrophysics, University of California, Santa Cruz, CA 95064, USA\\
$^{16}$ Department of Physics and Astronomy, University of New Mexico, 210 Yale Blvd NE, Albuquerque, NM 87106, USA\\
$^{17}$ Institute of Planetary Research, German Aerospace Center, Rutherfordstrasse 2, 12489 Berlin, Germany\\
$^{18}$ Max Planck Institute for Astronomy, K{\"{o}}nigstuhl 17, 69117 - Heidelberg, Germany\\
$^{19}$ Department of Physics and Kavli Institute for Astrophysics and Space Research, Massachusetts Institute of Technology, 77 Massachusetts Avenue, Cambridge, MA 02139, USA\\
$^{20}$ European Space Agency (ESA), European Space Research and Technology Centre (ESTEC), Keplerlaan 1, 2201 AZ Noordwijk, The Netherlands\\
$^{21}$ Astronomy Unit, Queen Mary University of London, Mile End Road, London E1 4NS, UK\\
$^{22}$ Astrophysics Group, Cavendish Laboratory, J.J. Thomson Avenue, Cambridge, CB3 0HE, UK\\
$^{23}$ NASA Goddard Space Flight Center, 8800 Greenbelt Rd, Greenbelt, MD 20771, USA\\
$^{24}$ Instituto de Astronom\'ia, Universidad Cat\'olica del Norte, Angamos 0610, 1270709, Antofagasta, Chile\\
$^{25}$ N\'ucleo de Astronom\'ia, Facultad de Ingenier\'ia y Ciencias, Universidad Diego Portales, Av. Ej\'ercito 441, Santiago, Chile\\
$^{26}$ Space Research Institute, Austrian Academy of Sciences, Schmiedlstraße 6, 8042 Graz, Austria\\
$^{27}$ Instituto de Astrof´ısica, Facultad de F´ısica, Pontificia Universidad Cat´olica de Chile\\
$^{28}$ European Southern Observatory (ESO), Karl-Schwarzschild-Str. 2, 85748 Garching bei M{\"u}nchen, Germany\\
$^{29}$ Center for Data Intensive and Time Domain Astronomy, Department of Physics and Astronomy, Michigan State University, East Lansing, MI 48824, USA\\
$^{30}$ Armagh Observatory and Planetarium, College Hill, Armagh, BT61 9DG, UK\\
$^{31}$ Department of Astronomy/Steward Observatory, The University of Arizona, 933 North Cherry Avenue, Tucson, AZ 85721, USA\\
$^{32}$ Department of Earth, Atmospheric and Planetary Sciences, Massachusetts Institute of Technology, Cambridge, MA 02139, USA\\
$^{33}$ Department of Aeronautics and Astronautics, MIT, 77 Massachusetts Avenue, Cambridge, MA 02139, USA\\
$^{34}$ NASA Ames Research Center, Moffett Field, CA 94035, USA\\
$^{35}$ Department of Astrophysical Sciences, 4 Ivy Lane, Princeton, University, Princeton, NJ 08540 US\\
$^{36}$ Department of Astronomy, Tsinghua University, Beijing 100084, China\\
$^{37}$ University of Southern Queensland, Centre for Astrophysics, West Street, Toowoomba, QLD 4350 Australia\\
$^{38}$ Department of Astronomy, Sofia University ``St Kliment Ohridski'', 5 James Bourchier Blvd, BG-1164 Sofia, Bulgari\\
$^{39}$ Space Research and Planetary Sciences, Physics Institute, University of Bern, Gesellschaftsstrasse 6, 3012 Bern, Switzerland 
\section{Transit timing variations}
\begin{figure*}
    \includegraphics[width=0.48\textwidth]{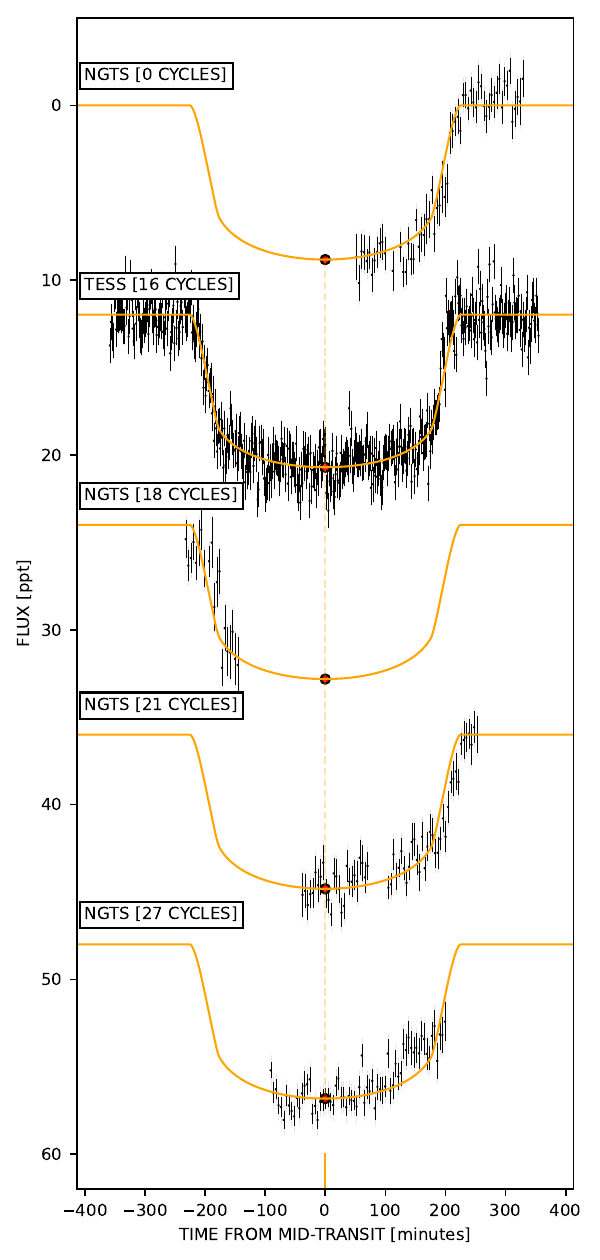}
    \includegraphics[width=0.48\textwidth]{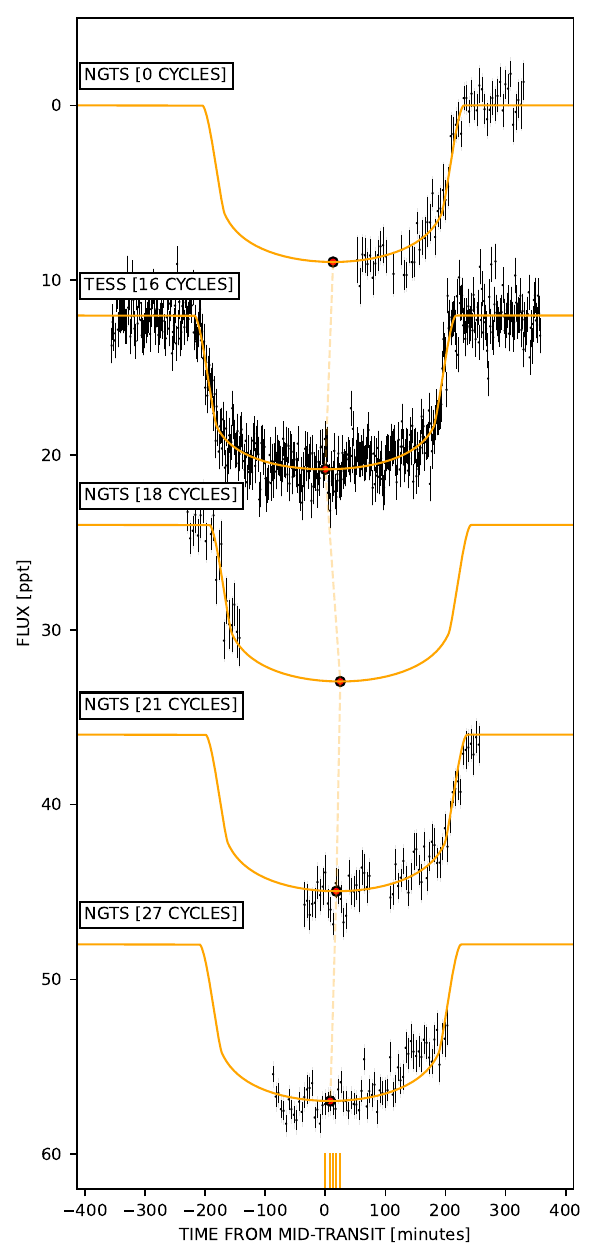}
    \caption{Modelling of \tess\ and \ngts\ transit events of \systemtb$\slash$\systemnb\ with transiting variations (right) and without (left). In each case, the fitted times of transit centre are marked for each event (red dot; joined with orange-dashed line) and marked with orange ticks on the lower axis.}
    \label{fig:appendix:ttv_justification}
\end{figure*}
\twocolumn

\bsp	
\label{lastpage}
\end{document}